\documentclass[12pt,showpacs,showkeys,superscriptaddress,aps]{revtex4}
\usepackage{amsmath,amsthm,amssymb}
\usepackage{epic, eepic}
\usepackage{graphicx} 
\newcommand{\maprightu}[1]{%
\smash{\mathop{%
\hbox to 1cm{\rightarrowfill}}\limits^{#1}}}
\newcommand{\maprightd}[1]{%
\smash{\mathop{%
\hbox to 1cm{\rightarrowfill}}\limits_{#1}}}
\newcommand{\mapleftu}[1]{%
\smash{\mathop{%
\hbox to 1cm{\leftarrowfill}}\limits^{#1}}}
\newcommand{\mapleftd}[1]{%
\smash{\mathop{%
\hbox to 1cm{\leftarrowfill}}\limits_{#1}}}
\newcommand{\mapnerssss}[1]{%
\smash{\mathop{%
\hbox to 3cm{\nearrow}}\limits^{#1}}}
\begin{document}
\title{Degeneration of the Julia set to singular loci of algebraic curves}
\author{Satoru Saito}
\email[email : ]{saito_ru@nifty.com}
\affiliation{Hakusan 4-19-10, Midori-ku, Yokohama 226-0006 Japan}
\author{Noriko Saitoh}
\email[email : ]{nsaitoh@ynu.ac.jp}
\affiliation{Department of Applied Mathematics, Yokohama National University\\
Hodogaya-ku, Yokohama, 240-8501 Japan}
\author{Hiromitsu Harada}
\email[email : ]{harada-hiromitsu@ed.tmu.ac.jp}
\affiliation{Department of Physics, Tokyo Metropolitan University,\\
Minamiohsawa 1-1, Hachiohji, Tokyo, 192-0397 Japan}
\author{Tsukasa YumibayashiI}
\email[email : ]{yumibayashi-tsukasa@ed.tmu.ac.jp}
\affiliation{Department of Physics, Tokyo Metropolitan University,\\
Minamiohsawa 1-1, Hachiohji, Tokyo, 192-0397 Japan}
\author{Yuki Wakimoto}
\email[email : ]{wakimoto-yuki@ed.tmu.ac.jp}
\affiliation{Department of Physics, Tokyo Metropolitan University,\\
Minamiohsawa 1-1, Hachiohji, Tokyo, 192-0397 Japan}
%%%%%%%%%%%%%%%%%%%%%%%%%%%%%%%%%%%%%%%
\keywords{ Julia set, singular locus, integrable limit}
%%%%%%%%%%%%%%%%%%%%%%%%%%%%%%%%%%%%%%%
\begin{abstract}
We show that, when a non-integrable rational map changes to an integrable one continuously, a large part of the Julia set of the map approach indeterminate points (IDP) of the map along algebraic curves. We will see that the IDPs are singular loci of the curves.
\end{abstract}

\maketitle

\section{Introduction}

When we study the transition of a non-integrable map to an integrable one, we must know the behavior of the Julia set \cite{Devaney}. If there exists the Julia set, some orbits generated by the map are disturbed by the  set and behave chaotic. On the other hand an integrable map has no Julia set, so that we can decide the behavior analytically for all initial points.

The Julia set is the closure of the set of repulsive periodic points of a map. It has been studied, for a long time, how it behaves and how it is created from integrable maps. There have been known some important results, such as the Poincar\'e-Birkhoff fixed point theorem and the KAM theorem \cite{Birkhoff, Reichl}, which describe how the transition takes place. Since this 
phenomenon is quite singular, however, all results so far known are based on perturbation.

We would like to know analytically what happens at the transition point. How and where the Julia set disappears in an integrable limit. 
To explore the phenomenon it is convenient to study a map with a parameter which interpolates between these two regimes. 
We have been studying many rational maps in this way and have shown how periodic points of low periods move around as the parameter varies and where they approach in the integrable limit \cite{fate}.\\

We have shown, in our previous works \cite{fate, SS}, that the periodic points of an integrable map form a variety different for each period. We called such a variety an invariant variety of periodic points, or IVPP, because the variety is determined by the invariants of the map alone. Some of the periodic points of a non-integrable map were shown to approach IVPPs in the integrable limit. An interesting fact found in \cite{fate} was that there are many other points which move, instead of IVPP, to indeterminate points (IDP) of the map. In other words the IDP, where the denominator and the numerator of the map vanish simultaneously, are the source of the Julia set, such that the Julia set is created as the map becomes non-integrable. 
\bigskip

The work of \cite{fate} was done by studying every path of a periodic point analytically as the value of the parameter changes, by using computer algebra. Although this method is sufficient to have information of the path one by one, it does not provide global information. In this note we would like to study the global feature of the transition. Namely we derive algebraic curves along which the periodic points approach the IDPs. We will show that a set of periodic points of each period are on one algebraic curve and move to the IDP simultaneously in the integrable limit. This means that the IDP is a singular locus of the curve. Since periodic points of all periods approach there altogether, a large part of the Julia set itself degenerate to the singular locus, if they do not approach IVPPs.

The singular locus of algebraic hypersurface has been one of the main subjects to study in mathematics \cite{Milnor}. When the characteristic is zero, like our case, the Hironaka theorem shows desingularization of the singular locus. It is interesting to mention that the singularity of an algebraic curve corresponds to the integrable limit of non-integrable rational maps. We can provide many examples which show this correspondence, although we present here only one of them. 

\section{Paths of periodic points toward integrable limit} 

We consider, in this note, only one simple map to clarify our argument. We can present many other examples to support our results, but they are more or less similar. The map we consider is the following:
\begin{equation}
(x,y)\to F_a(x,y)=\left(x{1-y\over 1-x-a},\ 
y{1-x\over 1-y}\right)
\label{F_a}
\end{equation}
where $x,y\in\mathbb{C}$ and $a\in\mathbb{C}$ is a continuous parameter. Notice that, when $a=0$, this map has an IDP at $(x,y)=(1,1)$.

\subsection{Integrable limit}

When $a=0$, the map $F_0(x,y)$ is integrable. In fact, $F_0(x,y)$ has the invariant
\[
r=xy
\]
and reduces to the M\"obius map
\[
x\to X= {x-r\over 1-x}.
\]

The periodic points of $n$ period of the map (\ref{F_a}) are obtained by solving the condition
\begin{equation}
F^{(n)}_a(x,y)=(x,y),\qquad n=2,3,4,....
\label{PP condition}
\end{equation}
When the map is integrable, however, there is a simple way to find them. Namely the singularity confinement enables us to generate IVPPs iteratively \cite{fate, YSW}.
To start with we choose an initial point at $p^{(0)}=(1,r)$, so that the map undergoes according to
\[
p^{(0)}\rightarrow (\infty,0)\rightarrow (-1,-r)\rightarrow \left(-{1+r\over 2},\ -{2r\over 1+r}\right)\qquad\qquad\qquad\qquad
\]
\begin{equation}
\rightarrow \left(-{1+3r\over 3+r},\ -{r(3+r)\over 1+3r}\right)\rightarrow
\left(-{1+6r+r^2\over 4(1+r)},\ -{4r(1+r)\over 1+6r+r^2}\right)\rightarrow
\cdots.
\label{Moebius}
\end{equation}
Since the point $p^{(n)}=(x^{(n)},y^{(n)})$, after $n$ steps of the map, must diverge at the periodic points of period $n-1$, the denominator of $x^{(n)}$ vanishes there. For example, from (\ref{Moebius}), we see that the periodic points of period 3 is on the line specified by $r+3=0$. In this way we find that the IVPP of period $n$ is given by the curve, 
\[
\Big\{ x,y \Big|\gamma^{(n)}(x,y)=0\Big\},\qquad
\gamma^{(n)}(x,y):=xy+\tan^2{\pi\over n},\qquad n=3,4,5,...
\]

%%%%%%%%%%%%%%%%%%%%%%%%%%%%%%%%%%%%%%%%%%%%%%%%%%%%%%
%%%%%%%%%%%%%%%%%%%%%%%%%%%%%%%%%%%%%

\unitlength 1pt
\begin{center}
\begin{picture}(207.4149,229.0959)( 28.9080,-244.2726)
% FUNC 2 0 3 0 Black White
% 9 400 400 3200 3200 1800 1800 2200 1800 1800 1400 400 400 3200 3200 0 2 0 0
% -3/x
\special{pn 8}%
\special{pa 400 1458}%
\special{pa 410 1456}%
\special{pa 416 1454}%
\special{pa 430 1450}%
\special{pa 436 1448}%
\special{pa 446 1446}%
\special{pa 450 1444}%
\special{pa 460 1442}%
\special{pa 466 1440}%
\special{pa 476 1438}%
\special{pa 480 1436}%
\special{pa 490 1434}%
\special{pa 496 1432}%
\special{pa 500 1432}%
\special{pa 506 1430}%
\special{pa 510 1428}%
\special{pa 516 1426}%
\special{pa 526 1424}%
\special{pa 530 1422}%
\special{pa 536 1422}%
\special{pa 540 1420}%
\special{pa 546 1418}%
\special{pa 556 1414}%
\special{pa 560 1414}%
\special{pa 566 1412}%
\special{pa 570 1410}%
\special{pa 576 1408}%
\special{pa 580 1408}%
\special{pa 590 1404}%
\special{pa 596 1402}%
\special{pa 606 1398}%
\special{pa 610 1398}%
\special{pa 626 1392}%
\special{pa 630 1390}%
\special{pa 646 1384}%
\special{pa 650 1384}%
\special{pa 680 1372}%
\special{pa 686 1370}%
\special{pa 720 1356}%
\special{pa 726 1354}%
\special{pa 756 1342}%
\special{pa 760 1338}%
\special{pa 776 1332}%
\special{pa 780 1330}%
\special{pa 790 1326}%
\special{pa 796 1322}%
\special{pa 806 1318}%
\special{pa 810 1316}%
\special{pa 816 1314}%
\special{pa 820 1310}%
\special{pa 826 1308}%
\special{pa 830 1306}%
\special{pa 836 1304}%
\special{pa 846 1298}%
\special{pa 850 1296}%
\special{pa 860 1290}%
\special{pa 866 1288}%
\special{pa 886 1276}%
\special{pa 890 1274}%
\special{pa 920 1256}%
\special{pa 926 1252}%
\special{pa 946 1240}%
\special{pa 950 1236}%
\special{pa 960 1230}%
\special{pa 966 1226}%
\special{pa 970 1222}%
\special{pa 976 1218}%
\special{pa 980 1216}%
\special{pa 990 1208}%
\special{pa 996 1204}%
\special{pa 1010 1192}%
\special{pa 1016 1190}%
\special{pa 1036 1174}%
\special{pa 1040 1168}%
\special{pa 1056 1156}%
\special{pa 1060 1152}%
\special{pa 1066 1148}%
\special{pa 1070 1142}%
\special{pa 1076 1138}%
\special{pa 1080 1134}%
\special{pa 1086 1130}%
\special{pa 1130 1084}%
\special{pa 1146 1068}%
\special{pa 1150 1062}%
\special{pa 1186 1020}%
\special{pa 1190 1014}%
\special{pa 1196 1008}%
\special{pa 1230 958}%
\special{pa 1236 950}%
\special{pa 1240 944}%
\special{pa 1266 904}%
\special{pa 1270 894}%
\special{pa 1276 886}%
\special{pa 1296 850}%
\special{pa 1320 800}%
\special{pa 1326 790}%
\special{pa 1330 780}%
\special{pa 1340 758}%
\special{pa 1366 698}%
\special{pa 1376 672}%
\special{pa 1396 616}%
\special{pa 1406 586}%
\special{pa 1420 538}%
\special{pa 1430 504}%
\special{pa 1440 468}%
\special{pa 1450 430}%
\special{pa 1456 410}%
\special{pa 1458 400}%
\special{fp}%
\special{pa 2144 3200}%
\special{pa 2146 3192}%
\special{pa 2150 3172}%
\special{pa 2160 3134}%
\special{pa 2170 3098}%
\special{pa 2180 3064}%
\special{pa 2206 2986}%
\special{pa 2226 2930}%
\special{pa 2236 2904}%
\special{pa 2260 2844}%
\special{pa 2270 2822}%
\special{pa 2276 2812}%
\special{pa 2280 2800}%
\special{pa 2306 2750}%
\special{pa 2326 2714}%
\special{pa 2330 2706}%
\special{pa 2336 2698}%
\special{pa 2360 2658}%
\special{pa 2366 2650}%
\special{pa 2370 2642}%
\special{pa 2406 2594}%
\special{pa 2410 2588}%
\special{pa 2416 2580}%
\special{pa 2450 2538}%
\special{pa 2456 2534}%
\special{pa 2460 2528}%
\special{pa 2466 2522}%
\special{pa 2470 2516}%
\special{pa 2516 2472}%
\special{pa 2530 2458}%
\special{pa 2536 2454}%
\special{pa 2546 2444}%
\special{pa 2560 2432}%
\special{pa 2566 2428}%
\special{pa 2586 2412}%
\special{pa 2590 2408}%
\special{pa 2606 2396}%
\special{pa 2610 2394}%
\special{pa 2620 2386}%
\special{pa 2626 2382}%
\special{pa 2630 2378}%
\special{pa 2636 2376}%
\special{pa 2640 2372}%
\special{pa 2650 2366}%
\special{pa 2656 2362}%
\special{pa 2676 2350}%
\special{pa 2680 2346}%
\special{pa 2710 2328}%
\special{pa 2716 2326}%
\special{pa 2736 2314}%
\special{pa 2740 2312}%
\special{pa 2750 2306}%
\special{pa 2756 2304}%
\special{pa 2766 2298}%
\special{pa 2770 2296}%
\special{pa 2776 2292}%
\special{pa 2780 2290}%
\special{pa 2786 2288}%
\special{pa 2790 2286}%
\special{pa 2796 2282}%
\special{pa 2806 2278}%
\special{pa 2810 2276}%
\special{pa 2820 2272}%
\special{pa 2826 2268}%
\special{pa 2840 2262}%
\special{pa 2846 2260}%
\special{pa 2876 2248}%
\special{pa 2880 2244}%
\special{pa 2916 2230}%
\special{pa 2920 2230}%
\special{pa 2950 2218}%
\special{pa 2956 2216}%
\special{pa 2970 2210}%
\special{pa 2976 2210}%
\special{pa 2990 2204}%
\special{pa 2996 2202}%
\special{pa 3006 2198}%
\special{pa 3010 2198}%
\special{pa 3020 2194}%
\special{pa 3026 2192}%
\special{pa 3030 2190}%
\special{pa 3036 2190}%
\special{pa 3040 2188}%
\special{pa 3046 2186}%
\special{pa 3056 2182}%
\special{pa 3060 2182}%
\special{pa 3066 2180}%
\special{pa 3070 2178}%
\special{pa 3076 2176}%
\special{pa 3086 2174}%
\special{pa 3090 2172}%
\special{pa 3096 2172}%
\special{pa 3100 2170}%
\special{pa 3106 2168}%
\special{pa 3110 2166}%
\special{pa 3120 2164}%
\special{pa 3126 2162}%
\special{pa 3136 2160}%
\special{pa 3140 2158}%
\special{pa 3150 2156}%
\special{pa 3156 2154}%
\special{pa 3166 2152}%
\special{pa 3170 2150}%
\special{pa 3186 2148}%
\special{pa 3190 2146}%
\special{pa 3200 2144}%
\special{fp}%
% FUNC 2 0 3 0 Black White
% 10 400 400 3200 3200 1800 1800 2200 1800 1800 1400 400 400 3200 3200 0 2 0 0 0 0
% -1/x
\special{pn 8}%
\special{pa 400 1686}%
\special{pa 410 1686}%
\special{pa 416 1684}%
\special{pa 436 1684}%
\special{pa 440 1682}%
\special{pa 460 1682}%
\special{pa 466 1680}%
\special{pa 480 1680}%
\special{pa 486 1678}%
\special{pa 500 1678}%
\special{pa 506 1676}%
\special{pa 526 1676}%
\special{pa 530 1674}%
\special{pa 546 1674}%
\special{pa 556 1672}%
\special{pa 560 1672}%
\special{pa 566 1670}%
\special{pa 580 1670}%
\special{pa 586 1668}%
\special{pa 600 1668}%
\special{pa 606 1666}%
\special{pa 610 1666}%
\special{pa 620 1664}%
\special{pa 636 1664}%
\special{pa 646 1662}%
\special{pa 650 1662}%
\special{pa 656 1660}%
\special{pa 660 1660}%
\special{pa 670 1658}%
\special{pa 686 1658}%
\special{pa 696 1656}%
\special{pa 700 1656}%
\special{pa 710 1654}%
\special{pa 716 1654}%
\special{pa 730 1650}%
\special{pa 736 1650}%
\special{pa 746 1648}%
\special{pa 750 1648}%
\special{pa 766 1646}%
\special{pa 770 1646}%
\special{pa 786 1642}%
\special{pa 790 1642}%
\special{pa 810 1638}%
\special{pa 816 1638}%
\special{pa 846 1632}%
\special{pa 850 1632}%
\special{pa 960 1610}%
\special{pa 966 1608}%
\special{pa 986 1604}%
\special{pa 990 1602}%
\special{pa 1006 1600}%
\special{pa 1010 1598}%
\special{pa 1026 1594}%
\special{pa 1030 1592}%
\special{pa 1036 1592}%
\special{pa 1040 1590}%
\special{pa 1050 1588}%
\special{pa 1056 1586}%
\special{pa 1060 1584}%
\special{pa 1066 1582}%
\special{pa 1070 1582}%
\special{pa 1076 1580}%
\special{pa 1080 1578}%
\special{pa 1086 1576}%
\special{pa 1090 1576}%
\special{pa 1100 1572}%
\special{pa 1106 1570}%
\special{pa 1116 1566}%
\special{pa 1120 1566}%
\special{pa 1136 1560}%
\special{pa 1140 1558}%
\special{pa 1196 1536}%
\special{pa 1200 1534}%
\special{pa 1210 1530}%
\special{pa 1216 1526}%
\special{pa 1226 1522}%
\special{pa 1230 1520}%
\special{pa 1236 1518}%
\special{pa 1240 1514}%
\special{pa 1246 1512}%
\special{pa 1256 1506}%
\special{pa 1260 1504}%
\special{pa 1306 1478}%
\special{pa 1310 1474}%
\special{pa 1320 1468}%
\special{pa 1326 1464}%
\special{pa 1330 1460}%
\special{pa 1376 1424}%
\special{pa 1426 1374}%
\special{pa 1430 1368}%
\special{pa 1440 1356}%
\special{pa 1446 1350}%
\special{pa 1450 1344}%
\special{pa 1476 1308}%
\special{pa 1490 1284}%
\special{pa 1496 1276}%
\special{pa 1500 1268}%
\special{pa 1506 1258}%
\special{pa 1510 1248}%
\special{pa 1516 1240}%
\special{pa 1520 1230}%
\special{pa 1540 1186}%
\special{pa 1546 1174}%
\special{pa 1556 1148}%
\special{pa 1566 1120}%
\special{pa 1576 1090}%
\special{pa 1580 1074}%
\special{pa 1586 1056}%
\special{pa 1596 1020}%
\special{pa 1600 1000}%
\special{pa 1610 958}%
\special{pa 1616 936}%
\special{pa 1620 912}%
\special{pa 1626 886}%
\special{pa 1630 860}%
\special{pa 1636 830}%
\special{pa 1640 800}%
\special{pa 1646 768}%
\special{pa 1650 734}%
\special{pa 1656 698}%
\special{pa 1660 658}%
\special{pa 1666 616}%
\special{pa 1670 570}%
\special{pa 1676 520}%
\special{pa 1680 468}%
\special{pa 1686 410}%
\special{pa 1686 400}%
\special{fp}%
\special{pa 1914 3200}%
\special{pa 1916 3192}%
\special{pa 1920 3134}%
\special{pa 1926 3080}%
\special{pa 1930 3032}%
\special{pa 1936 2986}%
\special{pa 1940 2944}%
\special{pa 1946 2904}%
\special{pa 1950 2868}%
\special{pa 1956 2832}%
\special{pa 1960 2800}%
\special{pa 1966 2770}%
\special{pa 1970 2742}%
\special{pa 1976 2714}%
\special{pa 1980 2690}%
\special{pa 1986 2666}%
\special{pa 1990 2642}%
\special{pa 2000 2600}%
\special{pa 2006 2580}%
\special{pa 2016 2544}%
\special{pa 2020 2528}%
\special{pa 2026 2512}%
\special{pa 2036 2482}%
\special{pa 2046 2454}%
\special{pa 2056 2428}%
\special{pa 2060 2416}%
\special{pa 2080 2372}%
\special{pa 2086 2362}%
\special{pa 2090 2352}%
\special{pa 2096 2342}%
\special{pa 2100 2334}%
\special{pa 2106 2326}%
\special{pa 2110 2316}%
\special{pa 2126 2292}%
\special{pa 2150 2258}%
\special{pa 2156 2252}%
\special{pa 2160 2244}%
\special{pa 2170 2232}%
\special{pa 2176 2228}%
\special{pa 2180 2222}%
\special{pa 2186 2216}%
\special{pa 2190 2210}%
\special{pa 2210 2190}%
\special{pa 2216 2186}%
\special{pa 2226 2176}%
\special{pa 2270 2140}%
\special{pa 2276 2138}%
\special{pa 2280 2134}%
\special{pa 2290 2128}%
\special{pa 2296 2124}%
\special{pa 2340 2096}%
\special{pa 2346 2094}%
\special{pa 2356 2088}%
\special{pa 2360 2086}%
\special{pa 2366 2084}%
\special{pa 2370 2082}%
\special{pa 2376 2078}%
\special{pa 2386 2074}%
\special{pa 2390 2072}%
\special{pa 2400 2068}%
\special{pa 2406 2064}%
\special{pa 2460 2042}%
\special{pa 2466 2042}%
\special{pa 2480 2036}%
\special{pa 2486 2034}%
\special{pa 2496 2030}%
\special{pa 2500 2030}%
\special{pa 2510 2026}%
\special{pa 2516 2024}%
\special{pa 2520 2022}%
\special{pa 2526 2022}%
\special{pa 2530 2020}%
\special{pa 2536 2018}%
\special{pa 2540 2016}%
\special{pa 2546 2016}%
\special{pa 2550 2014}%
\special{pa 2560 2012}%
\special{pa 2566 2010}%
\special{pa 2570 2008}%
\special{pa 2576 2006}%
\special{pa 2590 2004}%
\special{pa 2596 2002}%
\special{pa 2610 1998}%
\special{pa 2616 1996}%
\special{pa 2636 1992}%
\special{pa 2640 1990}%
\special{pa 2750 1968}%
\special{pa 2756 1968}%
\special{pa 2786 1962}%
\special{pa 2790 1962}%
\special{pa 2810 1958}%
\special{pa 2816 1958}%
\special{pa 2830 1956}%
\special{pa 2836 1956}%
\special{pa 2850 1952}%
\special{pa 2856 1952}%
\special{pa 2866 1950}%
\special{pa 2870 1950}%
\special{pa 2886 1948}%
\special{pa 2890 1948}%
\special{pa 2900 1946}%
\special{pa 2906 1946}%
\special{pa 2916 1944}%
\special{pa 2920 1944}%
\special{pa 2926 1942}%
\special{pa 2930 1942}%
\special{pa 2940 1940}%
\special{pa 2956 1940}%
\special{pa 2966 1938}%
\special{pa 2970 1938}%
\special{pa 2976 1936}%
\special{pa 2980 1936}%
\special{pa 2990 1934}%
\special{pa 3006 1934}%
\special{pa 3010 1932}%
\special{pa 3026 1932}%
\special{pa 3030 1930}%
\special{pa 3046 1930}%
\special{pa 3056 1928}%
\special{pa 3060 1928}%
\special{pa 3066 1926}%
\special{pa 3086 1926}%
\special{pa 3090 1924}%
\special{pa 3106 1924}%
\special{pa 3110 1922}%
\special{pa 3126 1922}%
\special{pa 3130 1920}%
\special{pa 3150 1920}%
\special{pa 3156 1918}%
\special{pa 3170 1918}%
\special{pa 3176 1916}%
\special{pa 3196 1916}%
\special{pa 3200 1914}%
\special{fp}%
% FUNC 2 0 3 0 Black White
% 10 400 400 3200 3200 1800 1800 2200 1800 1800 1400 400 400 3200 3200 0 2 0 0 0 0
% -0.527864/x
\special{pn 8}%
\special{pa 400 1740}%
\special{pa 426 1740}%
\special{pa 430 1738}%
\special{pa 466 1738}%
\special{pa 470 1736}%
\special{pa 510 1736}%
\special{pa 516 1734}%
\special{pa 546 1734}%
\special{pa 550 1732}%
\special{pa 580 1732}%
\special{pa 586 1730}%
\special{pa 616 1730}%
\special{pa 620 1728}%
\special{pa 650 1728}%
\special{pa 656 1726}%
\special{pa 680 1726}%
\special{pa 686 1724}%
\special{pa 710 1724}%
\special{pa 716 1722}%
\special{pa 736 1722}%
\special{pa 740 1720}%
\special{pa 760 1720}%
\special{pa 766 1718}%
\special{pa 786 1718}%
\special{pa 790 1716}%
\special{pa 810 1716}%
\special{pa 816 1714}%
\special{pa 830 1714}%
\special{pa 836 1712}%
\special{pa 856 1712}%
\special{pa 860 1710}%
\special{pa 876 1710}%
\special{pa 880 1708}%
\special{pa 896 1708}%
\special{pa 900 1706}%
\special{pa 916 1706}%
\special{pa 926 1704}%
\special{pa 930 1704}%
\special{pa 936 1702}%
\special{pa 950 1702}%
\special{pa 960 1700}%
\special{pa 966 1700}%
\special{pa 970 1698}%
\special{pa 976 1698}%
\special{pa 986 1696}%
\special{pa 990 1696}%
\special{pa 1000 1694}%
\special{pa 1006 1694}%
\special{pa 1016 1692}%
\special{pa 1020 1692}%
\special{pa 1030 1690}%
\special{pa 1036 1690}%
\special{pa 1050 1688}%
\special{pa 1056 1688}%
\special{pa 1070 1684}%
\special{pa 1076 1684}%
\special{pa 1106 1678}%
\special{pa 1110 1678}%
\special{pa 1190 1662}%
\special{pa 1196 1660}%
\special{pa 1216 1656}%
\special{pa 1220 1654}%
\special{pa 1236 1652}%
\special{pa 1240 1650}%
\special{pa 1246 1648}%
\special{pa 1250 1646}%
\special{pa 1260 1644}%
\special{pa 1266 1642}%
\special{pa 1270 1642}%
\special{pa 1276 1640}%
\special{pa 1280 1638}%
\special{pa 1290 1634}%
\special{pa 1296 1634}%
\special{pa 1306 1630}%
\special{pa 1310 1628}%
\special{pa 1370 1604}%
\special{pa 1376 1602}%
\special{pa 1380 1600}%
\special{pa 1386 1596}%
\special{pa 1390 1594}%
\special{pa 1396 1592}%
\special{pa 1400 1590}%
\special{pa 1410 1584}%
\special{pa 1416 1582}%
\special{pa 1436 1570}%
\special{pa 1440 1566}%
\special{pa 1450 1560}%
\special{pa 1456 1556}%
\special{pa 1460 1552}%
\special{pa 1490 1528}%
\special{pa 1500 1518}%
\special{pa 1506 1514}%
\special{pa 1516 1504}%
\special{pa 1520 1498}%
\special{pa 1526 1494}%
\special{pa 1546 1470}%
\special{pa 1566 1442}%
\special{pa 1576 1426}%
\special{pa 1600 1378}%
\special{pa 1606 1368}%
\special{pa 1620 1332}%
\special{pa 1630 1304}%
\special{pa 1636 1288}%
\special{pa 1640 1272}%
\special{pa 1646 1256}%
\special{pa 1650 1238}%
\special{pa 1656 1218}%
\special{pa 1660 1198}%
\special{pa 1666 1174}%
\special{pa 1670 1150}%
\special{pa 1676 1124}%
\special{pa 1680 1096}%
\special{pa 1686 1066}%
\special{pa 1690 1032}%
\special{pa 1696 996}%
\special{pa 1700 956}%
\special{pa 1706 912}%
\special{pa 1710 862}%
\special{pa 1716 806}%
\special{pa 1720 744}%
\special{pa 1726 674}%
\special{pa 1730 594}%
\special{pa 1736 502}%
\special{pa 1740 400}%
\special{fp}%
\special{pa 1860 3200}%
\special{pa 1866 3100}%
\special{pa 1870 3008}%
\special{pa 1876 2926}%
\special{pa 1880 2856}%
\special{pa 1886 2794}%
\special{pa 1890 2738}%
\special{pa 1896 2690}%
\special{pa 1900 2646}%
\special{pa 1906 2604}%
\special{pa 1910 2568}%
\special{pa 1916 2534}%
\special{pa 1920 2504}%
\special{pa 1926 2476}%
\special{pa 1930 2450}%
\special{pa 1936 2426}%
\special{pa 1940 2404}%
\special{pa 1946 2382}%
\special{pa 1950 2364}%
\special{pa 1956 2346}%
\special{pa 1960 2328}%
\special{pa 1966 2312}%
\special{pa 1970 2298}%
\special{pa 1980 2270}%
\special{pa 1996 2234}%
\special{pa 2000 2222}%
\special{pa 2010 2202}%
\special{pa 2026 2176}%
\special{pa 2036 2160}%
\special{pa 2056 2132}%
\special{pa 2076 2108}%
\special{pa 2080 2102}%
\special{pa 2110 2072}%
\special{pa 2140 2048}%
\special{pa 2146 2046}%
\special{pa 2150 2042}%
\special{pa 2160 2036}%
\special{pa 2166 2032}%
\special{pa 2186 2020}%
\special{pa 2190 2018}%
\special{pa 2200 2012}%
\special{pa 2206 2010}%
\special{pa 2210 2006}%
\special{pa 2216 2004}%
\special{pa 2220 2002}%
\special{pa 2226 2000}%
\special{pa 2230 1996}%
\special{pa 2290 1972}%
\special{pa 2296 1972}%
\special{pa 2306 1968}%
\special{pa 2310 1966}%
\special{pa 2320 1962}%
\special{pa 2326 1962}%
\special{pa 2330 1960}%
\special{pa 2336 1958}%
\special{pa 2340 1956}%
\special{pa 2350 1954}%
\special{pa 2356 1952}%
\special{pa 2360 1952}%
\special{pa 2366 1950}%
\special{pa 2380 1946}%
\special{pa 2386 1944}%
\special{pa 2406 1940}%
\special{pa 2410 1938}%
\special{pa 2490 1922}%
\special{pa 2496 1922}%
\special{pa 2526 1916}%
\special{pa 2530 1916}%
\special{pa 2546 1914}%
\special{pa 2550 1914}%
\special{pa 2566 1910}%
\special{pa 2570 1910}%
\special{pa 2580 1908}%
\special{pa 2586 1908}%
\special{pa 2596 1906}%
\special{pa 2600 1906}%
\special{pa 2610 1904}%
\special{pa 2616 1904}%
\special{pa 2626 1902}%
\special{pa 2640 1902}%
\special{pa 2650 1900}%
\special{pa 2656 1900}%
\special{pa 2660 1898}%
\special{pa 2676 1898}%
\special{pa 2686 1896}%
\special{pa 2690 1896}%
\special{pa 2696 1894}%
\special{pa 2710 1894}%
\special{pa 2716 1892}%
\special{pa 2730 1892}%
\special{pa 2736 1890}%
\special{pa 2750 1890}%
\special{pa 2756 1888}%
\special{pa 2776 1888}%
\special{pa 2780 1886}%
\special{pa 2796 1886}%
\special{pa 2800 1884}%
\special{pa 2820 1884}%
\special{pa 2826 1882}%
\special{pa 2846 1882}%
\special{pa 2850 1880}%
\special{pa 2876 1880}%
\special{pa 2880 1878}%
\special{pa 2900 1878}%
\special{pa 2906 1876}%
\special{pa 2930 1876}%
\special{pa 2936 1874}%
\special{pa 2960 1874}%
\special{pa 2966 1872}%
\special{pa 2996 1872}%
\special{pa 3000 1870}%
\special{pa 3030 1870}%
\special{pa 3036 1868}%
\special{pa 3070 1868}%
\special{pa 3076 1866}%
\special{pa 3106 1866}%
\special{pa 3110 1864}%
\special{pa 3150 1864}%
\special{pa 3156 1862}%
\special{pa 3196 1862}%
\special{pa 3200 1860}%
\special{fp}%
% FUNC 2 0 3 0 Black White
% 10 400 400 3200 3200 1800 1800 2200 1800 1800 1400 400 400 3200 3200 0 2 0 0 0 0
% -1/(3*x)
\special{pn 8}%
\special{pa 400 1762}%
\special{pa 446 1762}%
\special{pa 450 1760}%
\special{pa 510 1760}%
\special{pa 516 1758}%
\special{pa 570 1758}%
\special{pa 576 1756}%
\special{pa 626 1756}%
\special{pa 630 1754}%
\special{pa 676 1754}%
\special{pa 680 1752}%
\special{pa 720 1752}%
\special{pa 726 1750}%
\special{pa 760 1750}%
\special{pa 766 1748}%
\special{pa 800 1748}%
\special{pa 806 1746}%
\special{pa 836 1746}%
\special{pa 840 1744}%
\special{pa 870 1744}%
\special{pa 876 1742}%
\special{pa 900 1742}%
\special{pa 906 1740}%
\special{pa 930 1740}%
\special{pa 936 1738}%
\special{pa 960 1738}%
\special{pa 966 1736}%
\special{pa 986 1736}%
\special{pa 990 1734}%
\special{pa 1006 1734}%
\special{pa 1010 1732}%
\special{pa 1030 1732}%
\special{pa 1036 1730}%
\special{pa 1050 1730}%
\special{pa 1056 1728}%
\special{pa 1070 1728}%
\special{pa 1076 1726}%
\special{pa 1090 1726}%
\special{pa 1096 1724}%
\special{pa 1110 1724}%
\special{pa 1116 1722}%
\special{pa 1120 1722}%
\special{pa 1130 1720}%
\special{pa 1146 1720}%
\special{pa 1156 1718}%
\special{pa 1160 1718}%
\special{pa 1170 1716}%
\special{pa 1176 1716}%
\special{pa 1186 1714}%
\special{pa 1190 1714}%
\special{pa 1206 1710}%
\special{pa 1210 1710}%
\special{pa 1230 1706}%
\special{pa 1236 1706}%
\special{pa 1330 1688}%
\special{pa 1336 1686}%
\special{pa 1346 1684}%
\special{pa 1350 1682}%
\special{pa 1360 1680}%
\special{pa 1366 1678}%
\special{pa 1376 1676}%
\special{pa 1386 1672}%
\special{pa 1390 1670}%
\special{pa 1396 1668}%
\special{pa 1400 1668}%
\special{pa 1416 1662}%
\special{pa 1420 1660}%
\special{pa 1450 1648}%
\special{pa 1456 1646}%
\special{pa 1466 1642}%
\special{pa 1470 1638}%
\special{pa 1476 1636}%
\special{pa 1480 1634}%
\special{pa 1486 1632}%
\special{pa 1520 1610}%
\special{pa 1530 1602}%
\special{pa 1536 1600}%
\special{pa 1550 1588}%
\special{pa 1580 1558}%
\special{pa 1596 1540}%
\special{pa 1616 1512}%
\special{pa 1620 1504}%
\special{pa 1636 1478}%
\special{pa 1640 1468}%
\special{pa 1646 1456}%
\special{pa 1656 1432}%
\special{pa 1660 1420}%
\special{pa 1666 1406}%
\special{pa 1670 1390}%
\special{pa 1680 1356}%
\special{pa 1686 1336}%
\special{pa 1690 1316}%
\special{pa 1696 1292}%
\special{pa 1700 1268}%
\special{pa 1706 1240}%
\special{pa 1710 1208}%
\special{pa 1716 1174}%
\special{pa 1720 1134}%
\special{pa 1726 1090}%
\special{pa 1730 1038}%
\special{pa 1736 980}%
\special{pa 1740 912}%
\special{pa 1746 830}%
\special{pa 1750 734}%
\special{pa 1756 616}%
\special{pa 1760 468}%
\special{pa 1762 400}%
\special{fp}%
\special{pa 1838 3200}%
\special{pa 1840 3134}%
\special{pa 1846 2986}%
\special{pa 1850 2868}%
\special{pa 1856 2770}%
\special{pa 1860 2690}%
\special{pa 1866 2622}%
\special{pa 1870 2562}%
\special{pa 1876 2512}%
\special{pa 1880 2468}%
\special{pa 1886 2428}%
\special{pa 1890 2394}%
\special{pa 1896 2362}%
\special{pa 1900 2334}%
\special{pa 1906 2308}%
\special{pa 1910 2286}%
\special{pa 1916 2264}%
\special{pa 1920 2244}%
\special{pa 1930 2210}%
\special{pa 1936 2196}%
\special{pa 1940 2182}%
\special{pa 1946 2168}%
\special{pa 1956 2144}%
\special{pa 1960 2134}%
\special{pa 1966 2124}%
\special{pa 1980 2096}%
\special{pa 1986 2088}%
\special{pa 2006 2060}%
\special{pa 2020 2042}%
\special{pa 2040 2022}%
\special{pa 2046 2018}%
\special{pa 2050 2014}%
\special{pa 2066 2002}%
\special{pa 2070 1998}%
\special{pa 2080 1990}%
\special{pa 2116 1970}%
\special{pa 2120 1968}%
\special{pa 2126 1964}%
\special{pa 2130 1962}%
\special{pa 2136 1960}%
\special{pa 2146 1956}%
\special{pa 2150 1952}%
\special{pa 2180 1940}%
\special{pa 2186 1940}%
\special{pa 2200 1934}%
\special{pa 2206 1932}%
\special{pa 2210 1930}%
\special{pa 2216 1930}%
\special{pa 2226 1926}%
\special{pa 2236 1924}%
\special{pa 2240 1922}%
\special{pa 2250 1920}%
\special{pa 2256 1918}%
\special{pa 2266 1916}%
\special{pa 2270 1914}%
\special{pa 2366 1894}%
\special{pa 2370 1894}%
\special{pa 2390 1890}%
\special{pa 2396 1890}%
\special{pa 2410 1888}%
\special{pa 2416 1888}%
\special{pa 2426 1886}%
\special{pa 2430 1886}%
\special{pa 2440 1884}%
\special{pa 2446 1884}%
\special{pa 2456 1882}%
\special{pa 2460 1882}%
\special{pa 2466 1880}%
\special{pa 2470 1880}%
\special{pa 2480 1878}%
\special{pa 2496 1878}%
\special{pa 2500 1876}%
\special{pa 2516 1876}%
\special{pa 2520 1874}%
\special{pa 2536 1874}%
\special{pa 2540 1872}%
\special{pa 2556 1872}%
\special{pa 2560 1870}%
\special{pa 2576 1870}%
\special{pa 2580 1868}%
\special{pa 2600 1868}%
\special{pa 2606 1866}%
\special{pa 2626 1866}%
\special{pa 2630 1864}%
\special{pa 2650 1864}%
\special{pa 2656 1862}%
\special{pa 2680 1862}%
\special{pa 2686 1860}%
\special{pa 2710 1860}%
\special{pa 2716 1858}%
\special{pa 2740 1858}%
\special{pa 2746 1856}%
\special{pa 2776 1856}%
\special{pa 2780 1854}%
\special{pa 2816 1854}%
\special{pa 2820 1852}%
\special{pa 2856 1852}%
\special{pa 2860 1850}%
\special{pa 2896 1850}%
\special{pa 2900 1848}%
\special{pa 2946 1848}%
\special{pa 2950 1846}%
\special{pa 2996 1846}%
\special{pa 3000 1844}%
\special{pa 3050 1844}%
\special{pa 3056 1842}%
\special{pa 3116 1842}%
\special{pa 3120 1840}%
\special{pa 3186 1840}%
\special{pa 3190 1838}%
\special{pa 3200 1838}%
\special{fp}%
% FUNC 2 0 3 0 Black White
% 10 400 400 3200 3200 1800 1800 2200 1800 1800 1400 400 400 3200 3200 0 2 0 0 0 0
% -0.232/x
\special{pn 8}%
\special{pa 400 1774}%
\special{pa 450 1774}%
\special{pa 456 1772}%
\special{pa 540 1772}%
\special{pa 546 1770}%
\special{pa 620 1770}%
\special{pa 626 1768}%
\special{pa 690 1768}%
\special{pa 696 1766}%
\special{pa 750 1766}%
\special{pa 756 1764}%
\special{pa 810 1764}%
\special{pa 816 1762}%
\special{pa 860 1762}%
\special{pa 866 1760}%
\special{pa 906 1760}%
\special{pa 910 1758}%
\special{pa 946 1758}%
\special{pa 950 1756}%
\special{pa 980 1756}%
\special{pa 986 1754}%
\special{pa 1016 1754}%
\special{pa 1020 1752}%
\special{pa 1050 1752}%
\special{pa 1056 1750}%
\special{pa 1076 1750}%
\special{pa 1080 1748}%
\special{pa 1106 1748}%
\special{pa 1110 1746}%
\special{pa 1130 1746}%
\special{pa 1136 1744}%
\special{pa 1150 1744}%
\special{pa 1156 1742}%
\special{pa 1176 1742}%
\special{pa 1180 1740}%
\special{pa 1196 1740}%
\special{pa 1200 1738}%
\special{pa 1216 1738}%
\special{pa 1226 1736}%
\special{pa 1230 1736}%
\special{pa 1236 1734}%
\special{pa 1250 1734}%
\special{pa 1260 1732}%
\special{pa 1266 1732}%
\special{pa 1276 1730}%
\special{pa 1280 1730}%
\special{pa 1296 1726}%
\special{pa 1300 1726}%
\special{pa 1316 1724}%
\special{pa 1320 1724}%
\special{pa 1346 1718}%
\special{pa 1350 1718}%
\special{pa 1386 1712}%
\special{pa 1390 1710}%
\special{pa 1416 1704}%
\special{pa 1420 1702}%
\special{pa 1430 1700}%
\special{pa 1436 1698}%
\special{pa 1440 1698}%
\special{pa 1446 1696}%
\special{pa 1450 1694}%
\special{pa 1456 1692}%
\special{pa 1460 1692}%
\special{pa 1466 1690}%
\special{pa 1470 1688}%
\special{pa 1516 1670}%
\special{pa 1520 1668}%
\special{pa 1530 1664}%
\special{pa 1546 1654}%
\special{pa 1550 1652}%
\special{pa 1556 1648}%
\special{pa 1570 1640}%
\special{pa 1596 1620}%
\special{pa 1600 1614}%
\special{pa 1606 1610}%
\special{pa 1616 1600}%
\special{pa 1620 1594}%
\special{pa 1630 1582}%
\special{pa 1646 1562}%
\special{pa 1650 1554}%
\special{pa 1660 1536}%
\special{pa 1666 1526}%
\special{pa 1676 1504}%
\special{pa 1680 1492}%
\special{pa 1690 1464}%
\special{pa 1700 1430}%
\special{pa 1706 1410}%
\special{pa 1710 1388}%
\special{pa 1716 1364}%
\special{pa 1720 1336}%
\special{pa 1726 1306}%
\special{pa 1730 1270}%
\special{pa 1736 1230}%
\special{pa 1740 1182}%
\special{pa 1746 1126}%
\special{pa 1750 1058}%
\special{pa 1756 976}%
\special{pa 1760 872}%
\special{pa 1766 740}%
\special{pa 1770 564}%
\special{pa 1774 400}%
\special{fp}%
\special{pa 1828 3200}%
\special{pa 1830 3038}%
\special{pa 1836 2862}%
\special{pa 1840 2728}%
\special{pa 1846 2626}%
\special{pa 1850 2542}%
\special{pa 1856 2476}%
\special{pa 1860 2420}%
\special{pa 1866 2372}%
\special{pa 1870 2330}%
\special{pa 1876 2296}%
\special{pa 1880 2264}%
\special{pa 1886 2238}%
\special{pa 1890 2212}%
\special{pa 1896 2192}%
\special{pa 1900 2172}%
\special{pa 1910 2138}%
\special{pa 1920 2110}%
\special{pa 1926 2098}%
\special{pa 1936 2076}%
\special{pa 1940 2066}%
\special{pa 1950 2048}%
\special{pa 1956 2040}%
\special{pa 1970 2018}%
\special{pa 1980 2006}%
\special{pa 1986 2002}%
\special{pa 1990 1996}%
\special{pa 2000 1986}%
\special{pa 2006 1982}%
\special{pa 2030 1962}%
\special{pa 2046 1952}%
\special{pa 2050 1948}%
\special{pa 2056 1946}%
\special{pa 2070 1938}%
\special{pa 2080 1934}%
\special{pa 2086 1930}%
\special{pa 2130 1912}%
\special{pa 2136 1912}%
\special{pa 2140 1910}%
\special{pa 2146 1908}%
\special{pa 2150 1906}%
\special{pa 2156 1906}%
\special{pa 2160 1904}%
\special{pa 2166 1902}%
\special{pa 2170 1900}%
\special{pa 2180 1898}%
\special{pa 2186 1896}%
\special{pa 2210 1892}%
\special{pa 2216 1890}%
\special{pa 2250 1882}%
\special{pa 2256 1882}%
\special{pa 2280 1878}%
\special{pa 2286 1878}%
\special{pa 2300 1874}%
\special{pa 2306 1874}%
\special{pa 2320 1872}%
\special{pa 2326 1872}%
\special{pa 2336 1870}%
\special{pa 2340 1870}%
\special{pa 2350 1868}%
\special{pa 2356 1868}%
\special{pa 2360 1866}%
\special{pa 2376 1866}%
\special{pa 2386 1864}%
\special{pa 2390 1864}%
\special{pa 2396 1862}%
\special{pa 2410 1862}%
\special{pa 2416 1860}%
\special{pa 2430 1860}%
\special{pa 2436 1858}%
\special{pa 2456 1858}%
\special{pa 2460 1856}%
\special{pa 2480 1856}%
\special{pa 2486 1854}%
\special{pa 2506 1854}%
\special{pa 2510 1852}%
\special{pa 2536 1852}%
\special{pa 2540 1850}%
\special{pa 2566 1850}%
\special{pa 2570 1848}%
\special{pa 2596 1848}%
\special{pa 2600 1846}%
\special{pa 2630 1846}%
\special{pa 2636 1844}%
\special{pa 2670 1844}%
\special{pa 2676 1842}%
\special{pa 2716 1842}%
\special{pa 2720 1840}%
\special{pa 2760 1840}%
\special{pa 2766 1838}%
\special{pa 2816 1838}%
\special{pa 2820 1836}%
\special{pa 2876 1836}%
\special{pa 2880 1834}%
\special{pa 2940 1834}%
\special{pa 2946 1832}%
\special{pa 3016 1832}%
\special{pa 3020 1830}%
\special{pa 3100 1830}%
\special{pa 3106 1828}%
\special{pa 3200 1828}%
\special{fp}%
% LINE 2 0 3 0 Black White
% 2 400 1800 3200 1800
% 
\special{pn 8}%
\special{pa 400 1800}%
\special{pa 3200 1800}%
\special{fp}%
% LINE 2 0 3 0 Black White
% 2 1800 400 1800 3200
% 
\special{pn 8}%
\special{pa 1800 400}%
\special{pa 1800 3200}%
\special{fp}%
% STR 2 0 3 0 Black White
% 4 1690 1840 1690 1940 2 0 0 0
% 0
\put(122.1363,-140.2038){\makebox(0,0)[lb]{0}}%
% STR 2 0 3 0 Black White
% 4 3270 1750 3270 1850 2 0 0 0
% $x$
\put(236.3229,-133.6995){\makebox(0,0)[lb]{$x$}}%
% STR 2 0 3 0 Black White
% 4 1710 240 1710 340 2 0 0 0
% $y$
\put(123.5817,-24.5718){\makebox(0,0)[lb]{$y$}}%
% STR 2 0 3 0 Black White
% 4 2480 2540 2480 2640 2 0 0 0
% $\gamma^{(3)}$
\put(179.2296,-190.7928){\makebox(0,0)[lb]{$\gamma^{(3)}$}}%
% STR 2 0 3 0 Black White
% 4 2150 2330 2150 2430 2 0 0 0
% $\gamma^{(4)}$
\put(155.3805,-175.6161){\makebox(0,0)[lb]{$\gamma^{(4)}$}}%
% DOT 0 0 3 0 Black White
% 2 2220 1400 2210 1410
% 
\special{pn 4}%
\special{sh 1}%
\special{ar 2220 1400 14 14 0  6.28318530717959E+0000}%
\special{sh 1}%
\special{ar 2210 1410 14 14 0  6.28318530717959E+0000}%
% STR 2 0 3 0 Black White
% 4 2210 1230 2210 1330 2 0 0 0
% $(1,1)$
\put(159.7167,-96.1191){\makebox(0,0)[lb]{$(1,1)$}}%
% STR 2 0 3 0 Black White
% 4 950 960 950 1060 2 0 0 0
% $\gamma^{(3)}$
\put(68.6565,-76.6062){\makebox(0,0)[lb]{$\gamma^{(3)}$}}%
% STR 2 0 3 0 Black White
% 4 1250 1310 1250 1410 2 0 0 0
% $\gamma^{(4)}$
\put(90.3375,-101.9007){\makebox(0,0)[lb]{$\gamma^{(4)}$}}%
% DOT 2 0 3 0 Black White
% 4 1660 1630 1700 1680 1740 1720 1740 1720
% 
\special{pn 4}%
\special{sh 1}%
\special{ar 1660 1630 6 6 0  6.28318530717959E+0000}%
\special{sh 1}%
\special{ar 1700 1680 6 6 0  6.28318530717959E+0000}%
\special{sh 1}%
\special{ar 1740 1720 6 6 0  6.28318530717959E+0000}%
\special{sh 1}%
\special{ar 1740 1720 6 6 0  6.28318530717959E+0000}%
% DOT 2 0 3 0 Black White
% 4 1850 1870 1890 1920 1930 1960 1930 1960
% 
\special{pn 4}%
\special{sh 1}%
\special{ar 1850 1870 6 6 0  6.28318530717959E+0000}%
\special{sh 1}%
\special{ar 1890 1920 6 6 0  6.28318530717959E+0000}%
\special{sh 1}%
\special{ar 1930 1960 6 6 0  6.28318530717959E+0000}%
\special{sh 1}%
\special{ar 1930 1960 6 6 0  6.28318530717959E+0000}%
% STR 2 0 3 0 Black White
% 4 1010 3410 1010 3510 2 0 0 0
% IVPPs of the M\"obius map
\put(72.9927,-253.6677){\makebox(0,0)[lb]{D1.  IVPPs of the M\"obius map}}%
\end{picture}%
\end{center}
%%%%%%%%%%%%%%%%%%%%%%%%%%%%%%%%%%%%%%%%%%%%%
%%%%%%%%%%%%%%%%%%%%%%%%%%%%%%%%%%%%%%%%%%
\bigskip

We notice that there is no IVPP of period 2 when $a=0$. 
\subsection{Fixed points}

There are two fixed points of $F_a(x,y)$ at
\[
(x,y)=(0,0)\  {\rm and}\  (-a,0),
\]

Although the fixed points are not considered being periodic in general, they form the line $y=x$ in the integrable limit in this particular map.

\subsection{Period 2 points}

When $a\ne 0$, we must solve the periodicity condition (\ref{PP condition}) to find the period 2 points. They are at 
$(x,y)=(x^{(2)}_j,y^{(2)}_j),\ j=1,2$, where
\begin{eqnarray}
(x^{(2)}_1,y^{(2)}_1)&=&\left(1-{a\over 2}+\sqrt{{a(a^2-4)\over a-4}},
\ 1+{a(2-a)\over 4}-{a\over 2}\sqrt{{a(a^2-4)\over a-4}}\right),\nonumber\\
(x^{(2)}_2,y^{(2)}_2)&=&\left(1-{a\over 2}-\sqrt{{a(a^2-4)\over a-4}},
\ 1+{a(2-a)\over 4}+{a\over 2}\sqrt{{a(a^2-4)\over a-4}}\right).
\label{period 2}
\end{eqnarray}
Now we recall that, in the $a=0$ case, there is no IVPP of period 2, while the period 2 points of (\ref{period 2}) exist at $a=0$. But the point $(x,y)=(1,1)$, where the period 2 points of (\ref{period 2}) approach, is exactly the IDP of the map $F_0(x,y)$. Therefore all period 2 points approach the IDP in this case, and none of them move to IVPP, in the integrable limit.\\
 
The explicit expression of the points like (\ref{period 2}) is not easy to find as the period $n$ becomes large. It will be more convenient to present the polynomial function $K_a^{(n)}(x)$ from which we can derive $x_j^{(n)}$ by solving $K_a^{(n)}(x)=0$, and another polynomial function $L_a^{(n)}(x)$ so that $y_j^{(n)}$ is given by $y_j^{(n)}=L^{(n)}_a(x_j^{(n)})$.
In the period 2 case we obtain
\begin{eqnarray}
K_a^{(2)}(x)&=&(a-4)x^2+(a-2)(a-4)x-2(a-2)(a-1),\nonumber\\
L_a^{(2)}(x)&=&\left(1-{a\over 2}\right)(x+a),
\label{K^2,L^2}
\end{eqnarray}
from which we find (\ref{period 2}) immediately. This is the way we studied the behavior of periodic points in our previous papers \cite{fate, SS, YSW}. \\

Now we want to know the paths of the periodic points along which they move as $a$ changes and approaches in the limit  $a=0$. We can do it if we eliminate the parameter $a$ from $K_a^{(2)}(x)=0$ and $y-L_a^{(2)}(x)=0$ of (\ref{K^2,L^2}). The result we obtain is the algebraic curve $G^{(2)}(x,y)=0$, with
\begin{equation}
G^{(2)}(x,y)=(2-x)^2(1-y)^2-3(1-x)(1-y)(2-x)
+(1-x)^2(2-x+xy).
\label{G2}
\end{equation}

This curve certainly passes the IDP $(1,1)$, hence $G^{(2)}(1,1)=0$, as we can check easily. From (\ref{K^2,L^2}) we see that it corresponds the integrable limit $a=0$. To see the paths of period 2 points, we can draw the curve on the $(x,y)$ real plane. We find a curve in FIG. 1.\\

\begin{figure}[h]
\begin{center}
\includegraphics[scale=0.3]{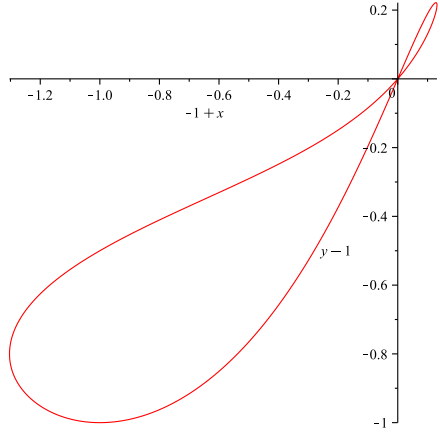}
\caption{\label{fig:sine}Path of period 2 points.}
\end{center}
\end{figure}

Notice that the IDP $(1,1)$ is shifted to the origin in this graph. 
From this picture it is apparent that the IDP is the singular locus of the curve (\ref{G2}). In fact we can convince ourselves that the multiplicity of the curve at the point $(1,1)$ is two. Since $K_a^{(2)}(x)$ and $L^{(2)}_a(x)$ are smooth functions of $a$, two periodic points must approach IDP along this curve at the same time, continuously as $a$ becomes small.

\subsection{Period 3 points}

By solving (\ref{PP condition}) for $n=3$, we obtain the functions $K_a^{(3)}(x)$ and $L_a^{(3)}(x)$, which are given by
\begin{eqnarray*}
K_a^{(3)}(x)&=&
6(a-2)x^9+3(a^2-13a+12+a^3)x^8+3(a-2)(3a^3-11a^2+11a-24)x^7\\&
+&(a-2)(6a^4-83a^3+289a^2-543a+456)x^6\\&
+&(516a^4-26a^5-2189a^3+4494a^2-7a^6-4686a+1944)x^5\\&
+&(124a^5-12a^7+6960a-8940a^2+5722a^3-1780a^4+66a^6-2088)x^4\\&
+&(63a^7+9335a^2-6a^8-5310a-8069a^3-142a^6-538a^5+3530a^4+1104)x^3\\&
-&(a-1)(a^2-3a+3)(a^6-16a^5+27a^4+170a^3-547a^2+462a-48)x^2\\&
+&(a^2-3a+3)(2a^5-9a^4-23a^3+108a^2-60a-36)(a-1)^2x\\&
-&(a^2-3a+3)(a^3+2a^2-9a-12)(a-1)^4\\&&\\
L_a^{(3)}(x)&=&{\begin{array}{l}
\Big[(a^2-3a+3)\Big(3x^4+2(4a-7)x^3+(a-2)(7a-8)x^2\cr
\qquad\qquad+(a-1)(2a^2-9a+6)x-(a-1)^3\Big)+2x^3(1-x)(a-2)\Big]\cr
\end{array}
\over
x(x+a-3)\Big(3x^3+3(2a-3)x^2+(3a^2-10a+5)x-2a^2+1+a\Big)}.
\end{eqnarray*}
From this result we see that there are nine period 3 points. It is already not easy to know how each of the nine points moves as the parameter $a$ changes. Nevertheless we can say that, as $a$ becomes small, three of them approach to the IVPP curve $xy+3=0$ and the rest move toward the IDP $(1,1)$. We can convince ourselves if we notice that $K_a^{(3)}(x)$ factorizes to
\[
K_0^{(3)}(x)=(x-1)^6(x^3+3x^2-9x-3)
\]
at $a=0$. \\

We have studied, in our previous work \cite{fate}, the behavior of the 3 point Lotka-Volterra map by using this method. The best we could do was to investigate the behavior of period 2 points. The new method we used in the previous subsection, however, suggests that we might get much more information if we can eliminate the parameter $a$ from $K_a^{(3)}(x)=0$ and $y-L_a^{(3)}(x)=0$. In fact we can use the elimination program of computer algebra to obtain an algebraic curve $G^{(3)}(x,y)=0$ in this way, where
\begin{eqnarray*}
G^{(3)}(x,y)&=&
x^3(x-1)^5y^9+x^2(3x^4-12x^3+15x^2-16x+15)(x-1)^3y^8\\&
-&x(x-1)(16x^7-81x^6+103x^5+109x^4-396x^3+392x^2-165x+18)y^7\\&
-&(-486+713x^3-13x^{10}+2394x^5+68x^9-1568x^6+1044x-220x^8\\&
+&717x^2-1864x^4+x^{11}+640x^7)y^6\\&
+&(-2059x^4+344x^9+5x^{11}+1278x+2493x^7+3570x^5-891-3542x^6\\&
-&46x^3-52x^2-1182x^8-58x^{10})y^5\\&
-&(-4025x^6+2493x^5-594-92x^{10}-497x^4+8x^{11}+3393x^7\\&
-&412x^2-462x^3+402x+498x^9-1656x^8)y^4\\&
+&(1436x^5+351x^9+150x^2-171+2129x^7+283x^4-2619x^6-1067x^8\\&
+&568x^3-24x-64x^{10}+4x^{11})y^3\\&
+&2(126x^3-3x^2-5x^4-68x^9-287x^5+9+8x^{10}+216x^8+429x^6\\&
+&374x^7+9x)y^2
+4x^3(47x^4-9+5x^6+15x^2-26x^5-38x^3-6x)y\\
&-&8x^6(-3x+x^2+3).
\end{eqnarray*}
If we plot the result we obtain a curve, as we show in Fig.2 and Fig.3.

%\begin{minipage}{8cm}
\begin{figure}[h]
\begin{center}
\includegraphics[scale=0.35]{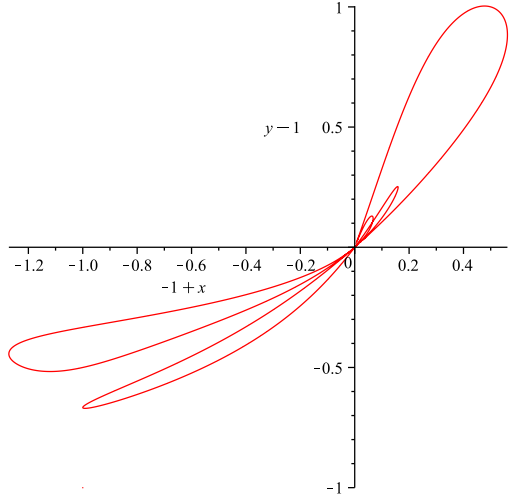}
\caption{Path of period 3 points.}
\end{center}
\end{figure}
%\end{minipage}
%\begin{minipage}{8cm}
\begin{figure}[h]
\includegraphics[scale=0.35]{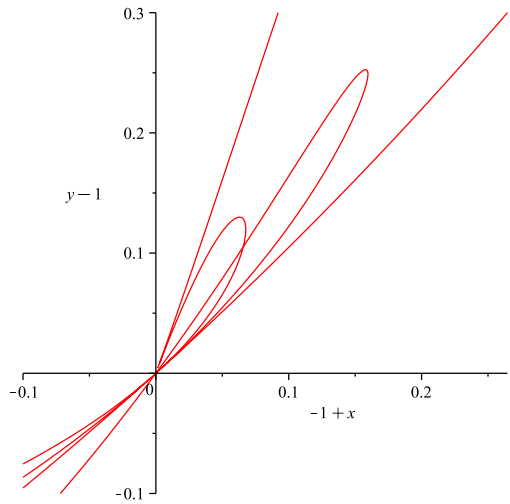}
\caption{Details of Fig.2 near IDP.}
\end{figure}
%\end{minipage}

This picture shows us many important information, which we can not guess from the algebraic expression of $G^{(3)}(x,y)$. The IDP $(1,1)$ is again the singular locus of the curve. The multiplicity of the point is six, so that six period 3 points approach the singularity in the integrable limit.

\subsection{Period 4 points}

We can use the computer algebra to solve the periodicity condition (\ref{PP condition}) in the case $n=4$ as well and obtain $K_a^{(4)}(x)$ and $L_a^{(4)}(x)$. Since their expressions are too big to present here, we report only the special case $a=0$ of $K_a^{(4)}(x)$. It is given by
\[
K_0^{(4)}(x)=(x-1)^{12}(x^4+4x^3-6x^2-4x+1)
\]
from which we can see that twelve out of sixteen points of period 4 move to the IDP $(1,1)$ and four other points approach the IVPP of period 4.

In order to find the paths of the periodic points as $a$ becomes small, we must derive the algebraic curve by the elimination of $a$ from $K_a^{(4)}(x)=0$ and $L_a^{(4)}(x)=y$. The formula of the curve $G^{(4)}(x,y)=0$, thus obtained, is the 27th degree in $x$ and the 22nd degree in $y$.  It is given in Appendix. If we draw the curve, we again confirm that the IDP $(1,1)$ is the singular locus of the curve. In the period 4 case we can check that this is a 12-ple point, hence twelve lines cross at this point, as we see in Fig.4 and Fig.5. 

%\begin{minipage}{8cm}
\begin{figure}[h]
\includegraphics[scale=0.35]{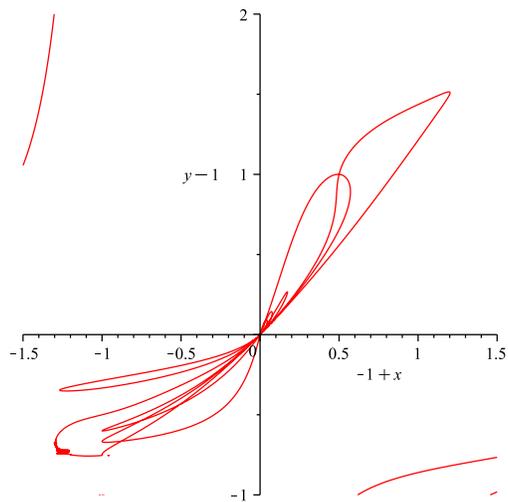}
\caption{Path of period 4 points.}
\end{figure}
%\end{minipage}
%\begin{minipage}{8cm}
\begin{figure}[h]
\includegraphics[scale=0.35]{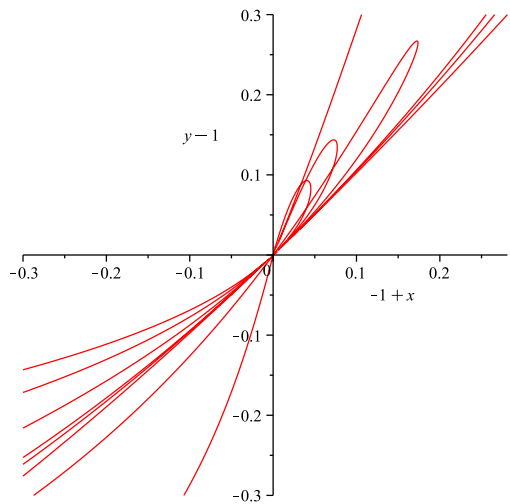}
\caption{Details of Fig.4 near IDP}
\end{figure}
%\end{minipage}

\section{Remarks}

Some remarks are in order.

\begin{enumerate}
\item
We have presented the behavior of periodic points in this note only those of small number of period.  Nevertheless we are certain that all other periodic points of higher periods behave similar, because of the IVPP theorem \cite{fate, SS}. In particular the periodic points will approach the IDP $(1,1)$ as $a$ becomes zero, if they do not move to the IVPPs. Since this is true for all periods, we can say that a large part of the Julia set approach the IDP. This is the phenomenon we found already in \cite{fate}, but we could not understand it well. Now the reason why this happens is clear because the IDP is a singular locus of the curve of each period. The loci of the curves of all periods degenerate there altogether. 

\item The Julia set is understood generally that it is produced via destruction of invariant tori by a perturbation. The Poincar\'e-Birkhoff fixed point theorem as well as the KAM theorem are based on this point of view \cite{Birkhoff, Reichl}. This corresponds to the part of the Julia set which approach to the IVPPs in our analysis. Since we know now that a large part of the Julia set is produced also from the IDPs of a rational map, we must change our view.

\item In this note we have studied a simple rational map, so that we can manage all formula explicitly. In particular the IDP of this map is a point. If we study higher dimensional maps, we should have a set of IDPs which might form a hypersurface. In fact the IDPs of $d$ dimensional Lotka-Volterra map form a variety of $d-2$ dimensional hypersurface. Therefore we should have higher dimensional singular locus of higher dimensional algebraic varieties. We will report some results in our forthcoming paper.
\end{enumerate}

\appendix

\section{Appendix}

The function $G^{(4)}(x,y)$ is as follows:
{\small
\begin{eqnarray*}
&&G^{(4)}(x,y)\\&&
=(8816x^{12}+256x^{10}-13857x^{17}-33909x^{15}-23x^{21}-2240x^{11}+5271x^{18}\\&&
+233x^{20}+25683x^{16}-1375x^{19}+x^{22}-20596x^{13}+31740x^{14})y^{22}
\\&&
+(490827x^{15}-396294x^{16}+26416x^{11}
-105480x^{18}-442001x^{14}-119x^{22}\\&&
+5x^{23}+1268x^{21}+237030x^{17}+34544x^{19}-114896x^{12}-8074x^{20}\\&&
-288x^9-1808x^{10}+278870x^{13})y^{21}
\\&&
+(2807x^{22}+11104x^8-2x^{25}-394x^{23}+273248x^{10}+1197014x^{12}-1963200x^{13}\\&&
-79856x^9+40x^{24}-633472x^{11}+2529495x^{16}-2971035x^{15}+803945x^{18}\\&&
-1642281x^{17}+80059x^{20}-16653x^{21}-293843x^{19}+2703024x^{14})y^{20}
\\
&&
+(18x^{25}+1695854x^{11}+92029x^{21}+88736x^8-388x^{24}+7472120x^{15}-21794x^{22}\\&&
+4259305x^{13}+4701916x^{17}+963340x^{19}-320012x^{20}+18408x^9-644680x^{10}\\&&
-6165979x^{14}-24640x^7-2409595x^{18}-6883117x^{16}-2825251x^{12}+3730x^{23})y^{19}
\\&&
+(3398840x^8-x^{26}+980318x^{19}-10510320x^{14}-5198624x^9\\&&
+11110019x^{13}-1214032x^7+809x^{24}+7506580x^{17}-158679x^{21}+191392x^6\\&&
-9693x^{23}-3809547x^{18}-8577042x^{12}+82868x^{20}+55769x^{22}\\&&
+10050359x^{15}-17x^{25}-9738850x^{16}+3600742x^{10}+2239109x^{11})y^{18}
\\&&
+(48136688x^{11}-46259480x^{12}-44009978x^{10}-104615268x^{17}-563680x^5-190x^{25}\\&&
+137427935x^{16}+29195734x^9+4343801x^{20}-112009280x^{15}+x^{27}-8926428x^8\\&&
+53115198x^{18}+23046203x^{13}+2481104x^6-11x^{26}+149650x^{22}+3914x^{24})y^{17}
\\&&
+(1092544x^4+188973x^{23}-20840938x^8-1204091x^{22}+57627698x^{19}\\&&
-173645515x^{11}+463350900x^{12}-5294816x^5+5013937x^{21}+10036456x^6\\&&
-1539x^{25}-157838037x^{18}-5479904x^7-14x^{27}+293x^{26}+60783280x^9\\&&
+226949466x^{15}+192896610x^{14}-17631661x^{20}-516416271x^{13}-10879x^{24}\\&&
+315090997x^{17}-402584676x^{16}-32082813x^{10})y^{16}
\\
\end{eqnarray*}
\begin{eqnarray*}
&&
+(-1950208x^3+6687808x^4-78187648x^6+151096242x^7-8407000x^{21}\\&&
+1566120x^{22}+55274x^{23}+605411253x^{10}-51206220x^{19}+118483207x^{18}\\&&
+30729419x^{15}-255344719x^{17}+327759324x^{16}-424750858x^9-1208014178x^{12}\\&&
+1593873742x^{13}+82x^{27}-2086x^{26}+21189x^{25}-96900x^{24}+7241576x^5\\&&
-4712192x^8+80889113x^{11}-914704386x^{14}+23562014x^{20})y^{15}
\\&&
+(2077184x^2+1119360x^3-35486112x^4+136255616x^{11}+1354268302x^{14}\\&&
+159769132x^6-492232314x^7-14803828x^{21}+39165117x^{20}+7459869x^{22}\\&&
-2786721x^{23}-1155871370x^{10}-145689877x^{19}+418399610x^{18}-1018324439x^{15}\\&&
-780021752x^{17}+972268702x^{16}+659067192x^9+1390921455x^{12}-1833303859x^{13}\\&&
-260x^{27}+7506x^{26}-93842x^{25}+656527x^{24}+45250844x^5+291928182x^8)y^{14}
\\
&&
+(313255516x^{11}-879722393x^{14}-839296x-11021120x^2-232431675x^{20}\\&&
+37576032x^3+14765908x^4+2160734x^6+462760612x^7+98525450x^{21}\\&&
-35954425x^{22}+9769283x^{23}-64253084x^{10}+558590205x^{19}-1322546200x^{18}\\&&
+2746049897x^{15}+2543750427x^{17}-3406782853x^{16}+144059212x^9+6579667x^{12}\\&&
-275687824x^{13}+481x^{27}-15606x^{26}+222017x^{25}-1831325x^{24}-142424100x^5\\&&
-564556180x^8)y^{13}
\\
&&
+(-4838400+17053952x+26423104x^2-163158392x^3+144722570x^4\\&&
-574867884x^6+385157548x^7-222319915x^{21}+512833376x^{20}+73412305x^{22}\\&&
-17505445x^{23}+2790214521x^{10}-997990618x^{19}+1868921028x^{18}\\&&
-4512279002x^{15}-3371530597x^{17}+4913319615x^{16}-2008832404x^9\\&&
-2312107559x^{12}+2500133310x^{13}-518x^{27}+19330x^{26}+1101010952x^{14}\\&&
+2890148x^{24}+232287632x^5+501944954x^8-884601685x^{11}-310710x^{25})y^{12}
\\&&
+(19825920-50415456x-62716320x^2-306833442x^4+1077001806x^6\\&&
-1220038452x^7+292546404x^{21}-704227113x^{20}-87855002x^{22}+330978924x^3\\&&
+18756682x^{23}-3476444777x^{10}+1250459787x^{19}-1764124791x^{18}+300x^{27}\\&&
+4468371630x^{15}+2400830340x^{17}-3556185396x^{16}+2922297534x^9\\&&
+2325948782x^{12}-646373274x^{13}-13845x^{26}+260022x^{25}-3021533059x^{14}\\&&
-2753015x^{24}-294473958x^5-273678892x^8+360386133x^{11})y^{11}
\\
\end{eqnarray*}
\begin{eqnarray*}
&&
+(-35939904+64769184x+115553224x^2-378347964x^3+266721942x^4\\&&
+1146474658x^7-245870933x^{21}+649653470x^{20}+65910006x^{22}-12401664x^{23}\\&&
+1033130062x^{10}-1189766722x^{19}+1416378080x^{18}-1843803742x^{15}\\&&
-986754541x^{17}+669723226x^{16}-1246384040x^9+1545608278x^{12}\\&&
-3868999655x^{13}-72x^{27}+5152x^{26}-124290x^{25}+1574659x^{24}-922292970x^6\\&&
+262022404x^5-142305796x^8-215331583x^{11}+3850801995x^{14})y^{10}
\\&&
+(37974512-41989448x-131341496x^2+251838780x^3-1956260354x^{14}\\&&
-101728022x^4+325964322x^6-552148272x^7+134646500x^{21}-404523573x^{20}\\&&
-31332231x^{22}-514714x^{24}+4998235x^{23}+1017113492x^{10}+829836329x^{19}\\&&
-1036638499x^{18}-858876830x^{15}+402699508x^{17}+861801359x^{16}\\&&
-1329186418x^9-5306469228x^{12}+5385746848x^{13}-732x^{26}+30148x^{25}\\&&
-112441894x^5+820000812x^8+1790817058x^{11})y^9
\\&&
+(-25913976+10793432x+90539488x^2-121110249x^{14}\\&&
-91478756x^3+942164x^4+103915466x^6+186707816x^7-48061450x^{21}\\&&
+169140349x^{20}+9315454x^{22}-1176908x^{23}-727043350x^{10}\\&&
-404422803x^{19}+600405225x^{18}+1437591457x^{15}-318328630x^{17}\\&&
-679751531x^{16}+2176833146x^9+5093465542x^{12}-3191570706x^{13}\\&&
-2804x^{25}+86768x^{24}-30060728x^5-1246953650x^8-2993945246x^{11})y^8
\\
&&
+(11940584+3293712x-37952592x^2+11302896x^3+12930744x^4-178584834x^6\\&&
-123180198x^7+10936264x^{21}-47103882x^{20}-1641007x^{22}+144020x^{23}\\&&
-176324386x^{10}+602080461x^{14}+134914901x^{19}-245425110x^{18}-755767973x^{15}\\&&
+209205466x^{17}+189135314x^{16}-1348734748x^9-2575626848x^{12}\\&&
+898925588x^{13}-5596x^{24}+70801624x^5+1024443616x^8+2310495792x^{11})y^7
\\
\end{eqnarray*}
\begin{eqnarray*}
&&
+(3862248x^3-84294803x^{17}+104166188x^7-4489028x^4+755984650x^{12}\\&&
-3764504+68848621x^{18}-301032512x^{14}-503516084x^8+359635802x^{10}\\&&
-30361454x^{19}+95239240x^6-6608x^{23}+148260x^{22}-1470444x^{21}\\&&
+212451793x^{15}-997934348x^{11}-64770610x^{13}-6521550x^{16}+8424977x^{20}\\&&
-45955316x^5+429526986x^9+9186216x^2-3666424x)y^6
\\&&
+(259154032x^{11}-165956678x^{10}+97348x^{21}+17621652x^5-4748x^{22}+78297118x^{14}\\&&
-991048x^2-34967968x^{15}+802280-62009120x^9-1714808x^3-132764454x^{12}\\&&
+21896808x^{17}-30379682x^{13}+831356x^4+1345800x-870582x^{20}+4357898x^{19}\\&&
+150628008x^8-13023482x^{18}-10925854x^{16}-53565568x^7-27540100x^6)y^5
\\&&
+(40960x^{20}-4471024x^5-51360x^2-26776688x^8-1996x^{21}+3888022x^{16}\\&&
+10683404x^{13}+14490878x^{12}-3699380x^{17}-209152x^4-578048x^9-110672\\&&
-263920x-13076026x^{14}-42943952x^{11}+4159888x^6+227600x^3+2813736x^{15}\\&&
-344788x^{19}+15613788x^7+1527780x^{18}+38852534x^{10})y^4
\\&&
+(5009038x^{11}-18294272x^{19}-29513x^{23}+39692303x^{14}-794239x^{21}-2090192x^7\\&&
-2507336x^7-897118x^{12}+2726588x^8+8928+1249024x^9-90772x^{18}+10864x^{19}\\&&
-5415958x^{10}+189702x^{15}-1959908x^{13}+27616x+361440x^{17}-163728x^6\\&&
-3648x^3+23872x^2-444x^{20}-675918x^{16}+56800x^4+745312x^5+1418766x^{14})y^3
\\&&
+(529396x^{10}-7552x^4-73280x^5-132528x^9+56484x^{16}-36392x^{12}\\&&
-320-63464x^{14}-15380x^{17}+227776x^{13}+1692x^{18}-40x^{19}+195920x^7\\&&
-72388x^{15}-1024x^3-425036x^{11}-1216x-35520x^6-182424x^8-1728x^2)y^2
\\&&
+(8608x^9+18824x^{11}+3136x^5 +12448x^{12}-1568x^{16}-12056x^{13}+320x^4\\&&
-4960x^7+5424x^{15}+14288x^8+3584x^6+120x^{17}-3088x^{14}-37848x^{10})y
\\&&
+1200x^{10}-80x^{11}-80x^{15}-640x^9-560x^{12}+80x^{13}-800x^8+240x^{14}
\end{eqnarray*}
}

\end{document}